\documentclass[prl,twocolumn,showkeys,showpacs]{revtex4}

\usepackage{epsfig}

\begin{document}

\title{Lattice Gas Dynamics; Application to Driven Vortices in Two Dimensional 
Superconductors}

\author{Violeta Gotcheva}
\affiliation{Department of Physics and Astronomy, University of
Rochester, Rochester, NY 14627}
\author{Albert T.~J.~Wang}
\altaffiliation{present address: Department of Physics, MIT}
\author{S. Teitel}
\affiliation{Department of Physics and Astronomy, University of
Rochester, Rochester, NY 14627}
\date{\today}

\begin{abstract}
A continuous time Monte Carlo lattice gas dynamics is developed
to model driven steady states of vortices in two dimensional superconducting 
networks.  Dramatic differences are found when compared to a simpler Metropolis
dynamics.  Subtle finite size effects are found at low temperature, with a moving
smectic that becomes unstable to an anisotropic liquid on sufficiently large length scales.
\end{abstract}
\maketitle


The nature of the driven steady states of 
many interacting particles, and the transitions between them, is a topic 
of much active interest.  As with equilibrium systems, the use of lattice models, 
in which the degrees of freedom sit on the sites of a discrete grid,
has led to analytical simplifications and greater accuracy in numerical simulations, 
as compared to continuum models.
\cite{Schmittmann}.  Here we consider lattice models applied to
driven two dimensional (2D) charges on a triangular grid, as 
a model for vortices in a honeycomb superconducting network.
We use two distinct lattice gas 
dynamics, both intended to model the overdamped diffusive limit: 
(i) the commonly used driven diffusive Metropolis Monte 
Carlo (MC) \cite{DDMC}, and its modification to a (ii) driven diffusive
continuous time Monte Carlo  \cite{Bortz, Newman}.  
We believe this is the first application of continuous time MC
in the context of driven diffusive problems.
We find that the steady states are qualitatively different 
for the two dynamics, and that the latter gives the more physically 
reasonable results.  We find that finite size effects can be
subtle at low temperature.

Our model is given by the Hamiltonian \cite{Franz},
\begin{equation}
    {\cal H}={1\over 2}\sum_{i,j}(n_{i}-f)V({\bf r}_{i}-{\bf r}_{j})
            (n_{j}-f)\enspace,
\label{eH}
\end{equation}
where $n_{i}=0,1$ is the charge on site ${\bf r}_{i}$ of a periodic 
triangular grid, $-f$ is a fixed uniform background charge, and 
$V({\bf r})$ is the 2D lattice Coulomb interaction as defined 
in Ref.\cite{Franz} for a triangular grid with periodic boundary 
conditions.  We take as grid basis vectors $\hat a_{1}=\hat x$,
and $\hat a_{2}={1\over 2}\hat x+{\sqrt{3}\over 2}\hat y$, the grid size 
is $L_{i}$ in direction $\hat a_{i}$, and the grid sites are
${\bf r}=m_{1}\hat a_{1}+m_{2}\hat a_{2}$,
$m_{i}=0,\ldots,L_{i}-1$.
Neutrality requires a fixed number of charges,  
$\sum_{i}n_{i}\equiv N_{c}=fL_{1}L_{2}$.
In this work we use a 
charge density of $f=1/25$.  In equilibrium, this model is 
characterized by a single first order melting transition at $T_{\rm m}\simeq 0.009$ 
from a triangular commensurate pinned solid with long range translational 
order to an ordinary liquid \cite{Franz}.

We now consider behavior in a uniform driving force, ${\bf 
F}=F\hat x$, parallel to the grid direction $\hat a_{1}$.
We consider two different dynamics, both involving single 
particle moves only.

(i) Driven diffusive Metropolis Monte Carlo (DDMC) \cite{DDMC}: At each step of the 
simulation, a charge $n_{i}=1$ is selected at random, and moved 
a distance $\Delta{\bf r}$ to a nearest neighbor site.  If $\hat 
a_{3}\equiv \hat a_{1}-\hat a_{2}$, then $\Delta{\bf r}$ is chosen 
randomly from the six possibilities, $\pm\hat a_{i}$, $i=1,2,3$. If ${\cal 
H}_{\rm old}$ and ${\cal H}_{\rm new}$ give the interaction energy (\ref{eH}) before  
and after the move, one computes,
\begin{equation}
    \Delta E\equiv {\cal H}_{\rm new}-{\cal H}_{\rm old}-{\bf 
    F}\cdot\Delta{\bf  r}\enspace,
\label{eDE}
\end{equation}
where the last term is the work done by the force on the moving charge.
One accepts or rejects this move according to the usual 
Metropolis MC algorithm.
One pass of $N_{c}$ steps equals one unit of simulation time.
Statistical averages are computed averaging over the generated 
configurations as in ordinary MC.  

(ii) Driven diffusive continuous time Monte Carlo (CTMC) \cite{Bortz, Newman}: 
At each step of the simulation, one considers the possible move of each charge $n_{i}=1$ in 
each of the six directions, $\hat\alpha=\pm\hat a_{i}$,
computing the energy change $\Delta E_{i\alpha}$ of each move 
according to Eq.(\ref{eDE}).  We take the rate for a particular 
move $i\alpha$ to be,
\begin{equation}
    W_{i\alpha}\equiv {W_{0}\rm e}^{-\Delta 
    E_{i\alpha}/2T}\enspace,
\label{eW}
\end{equation}
where $1/W_{0}$ sets the unit of time.
The total rate for all single particle moves is then
$W_{\rm tot}=\sum_{i\alpha}W_{i\alpha}$.
We decide which move to make by sampling the probability distribution 
$P_{i\alpha}\equiv W_{i\alpha}/W_{\rm tot}$, and then update the simulation clock by 
$\Delta t=1/W_{\rm tot}$.  Averages of an observable ${\cal O}$ are 
computed as,
\begin{equation}
    \langle{\cal O}\rangle = {1\over \tau}\int {\cal O}(t)dt
    ={1\over \tau}\sum_{s}{\cal O}_{s}\Delta t_{s}
\label{eO}
\end{equation}
where $s$ labels the steps of the simulation,
${\cal O}_{s}$ is the value of ${\cal O}$ in the configuration 
at time $t_{s}$, $\Delta t_{s}\equiv t_{s+1}-t_{s}$, and $\tau 
\equiv \sum_{s}\Delta t_{s}$ is the total time of the simulation.

The CTMC, originally introduced as the ``n-fold way'' for spin models \cite{Bortz}, 
owes its name to the continuous variations in the time 
steps $\Delta t_s$, which vary as the configuration changes throughout the
simulation. It is a {\it rejectionless} algorithm designed 
to speed up 
excitation over energy barriers at low temperatures; rather than 
waste many rejected moves until a rare acceptance takes one up 
an energy barrier, the energy barriers $\Delta E$ 
themselves set the time scale for each move, which then happens in a 
single step.  Simulation clock times can vary over orders of 
magnitude as $T$ varies.  

In CTMC, there are many possible choices for the rates that will obey local
detailed balance.  It can be shown \cite{Teitel} that the rates of Eq.(\ref{eW}) 
lead to ordinary Langevin dynamics in the limit $\Delta E_{i\alpha}/T<<1$.
Our simulations, however, are generally in the opposite limit
$\Delta E_{i\alpha}/T\agt 1$.  To see what physical limit this 
corresponds to, consider a single particle on a one 
dimensional grid in a driving 
force $F$. A simple calculation gives for the average velocity of 
the particle, $\langle v\rangle = W_{\rm tot}\tanh(F/2T)=2W_0\sinh(F/2T)$.
If we interpret the grid sites as the minima of a periodic pinning 
potential $U({\bf r})$ in the continuum, with energy barrier $U_{0}$, then $W_{0}\sim{\rm 
e}^{-U_{0}/T}$ and the above velocity then describes motion in such a 
periodic potential in the limit $F\alt U_{0}$ \cite{Ambegaokar}. CTMC thus 
appears to describe
the limit where motion is due to thermal activation over barriers; it is unclear
if it can describe the very large drive 
limit, where the washboard potential $U({\bf r})-{\bf F}\cdot{\bf r}$ 
loses its local minima parallel to ${\bf F}$.  This large drive limit
has been the subject of numerous theoretical \cite{Giamarchi, Balents, Scheidl} and numerical 
\cite{Faleski, Ryu, Dominguez, Fangohr} works 
for the case of random pinning.

In this paper we consider just the structural properties
of the steady state.  These are given by the structure function,

%
\begin{equation}
    S({\bf k})\equiv {1\over N_{c}}\langle q_{\bf k}q_{-{\bf k}}\rangle\enspace,
\label{eS}
\end{equation}
where $q_{\bf k}=\sum_{i}{\rm e}^{i{\bf k}\cdot{\bf r}_{i}}(n_{i}-f)$ is the 
Fourier transform of the charge distribution, ${\bf k}=k_{1}{\bf 
b}_{1}+k_{2}{\bf b}_{2}$ are the allowed wave vectors, with ${\bf
b}_{1}=2\pi\hat x-{2\pi\over\sqrt 3}\hat y$ and ${\bf
b}_{2}={4\pi\over\sqrt 3}\hat y$ the basis vectors of the
reciprocal lattice to the grid, and $k_{i}=l_{i}/L_{i}$ with
$l_{i}=0,\ldots L_{i-1}$.  We also consider the real space correlations,
\begin{equation}
    C(m_{1},k_{2})\equiv {1\over L_{1}}\sum_{k_{1}}{\rm
    e}^{-ik_{1}m_{1}}S(k_{1},k_{2})
\label{eC1}
\end{equation}
as well as $C(k_{1},m_{2})$ defined similarly, and  the 
6-fold (hexatic) orientational order parameter $\langle\Phi_{6}\rangle$, where
\begin{equation}
    \Phi_{6}\equiv {1\over N_{c}}\sum_{i}{1\over z_{i}}\sum_{j}{\rm 
    e}^{6i\theta_{ij}}\enspace.
\label{ePhi6}
\end{equation}
In the above, the first sum is over all charges $n_{i}=1$, the second 
sum is over all charges $n_{j}=1$ that are nearest neighbors of $n_{i}$
(as determined by a Delaunay triangulation), $z_{i}$ is the number of 
these nearest neighbors, and $\theta_{ij}$ is the angle of the bond 
from $n_{i}$ to $n_{j}$ with respect to the $\hat x$ axis.

The direct computation of $S({\bf k})$ and $\langle\Phi_{6}\rangle$
as in Eq.(\ref{eO}) is too costly as it involves lengthy calculations 
at each step of the simulation.  Instead we approximate the integral in 
Eq.(\ref{eO}) by a Monte Carlo evaluation, choosing $N_{\rm 
config}\simeq 1000$ random times uniformly distributed over the 
simulation interval 
$t\in [0,\tau]$ and averaging over the configurations at these times
only.  

\begin{figure}
\epsfxsize=7.5truecm
\epsfbox{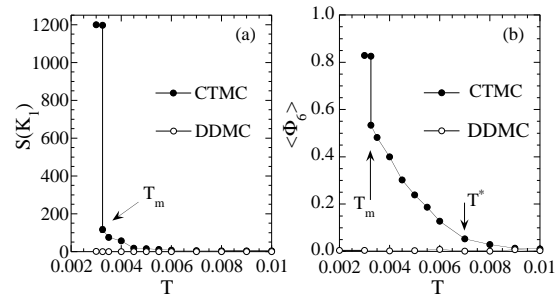}
\caption{(a) Plot of structure function transverse peak height 
$S({\bf K}_{1})$, ${\bf K}_{1}={1\over 5}{\bf
b}_{2}$, vs. $T$, and (b) hexatic order parameter
$\langle\Phi_{6}\rangle$ vs. $T$, at fixed ${\bf F}=0.1\hat x$, for both DDMC and
CTMC algorithms.
}
\label{f1}
\end{figure}
\begin{figure}
\epsfxsize=7.5truecm
\epsfbox{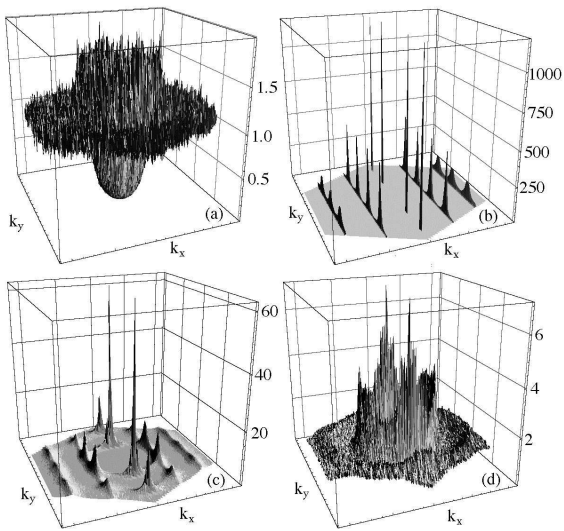}
\caption{Structure function $S({\bf k})$, for ${\bf k}$ in the first 
Brillouin zone, at force ${\bf F}=0.1\hat x$ for (a) $T=0.003$ with
DDMC algorithm; and (b) $T=0.003$, (c) $T=0.004$, (d) $T=0.007$ with 
CTMC algorithm.
}
\label{f2}
\end{figure}

We now present our results.  Starting in the ground state at $T=0$, 
the charge lattice remains pinned until the driving force
$F$ exceeds the change in interaction energy associated with moving
one charge forward one grid space parallel to ${\bf F}$.  This 
defines the $T=0$ critical force, $F_{c}=0.063$,  for both DDMC and 
CTMC.  Our simulations are carried out for fixed $F=0.1$ as 
we vary $T$.  Our results reported here are for a system size 
of $L_{1}=500$ and $L_{2}=60$, with $N_{c}=1200$ charges.  
The reason for such an extreme aspect 
ratio will be discussed later.  At $F=0$, $T<T_{\rm m}$, the system 
forms an ordered triangular charge solid, and $S({\bf k})$ has 
sharp Bragg peaks at reciprocal lattice vectors $\{{\bf K}\}$.
Let ${\bf K}_{1}={\bf b}_2/5$ be smallest non-zero reciprocal lattice 
vector directed transverse to ${\bf F}$.  In Fig.\,\ref{f1}a
we plot $S({\bf K}_{1})$ vs. $T$, at $F=0.1$, for both DDMC and 
CTMC.  In Fig.\,\ref{f1}b we plot $\langle\Phi_{6}\rangle$ 
vs. $T$.  
The results for DDMC show {\it no structure whatever} for the 
moving steady state.  A plot  in Fig.\,\ref{f2}a of the full $S({\bf 
k})$ at $T=0.003$ shows an isotropic liquid.  In fact, we find 
with DDMC that once the charged solid depins from the grid, the
moving steady state is an isotropic liquid virtually 
everywhere in the $F-T$ plane.

To see why this is so, consider a 
large $F>>F_{c}$ at $T=0$, starting from an ordered solid.  The DDMC 
picks a charge at random, then picks a direction to move it in at 
random; the move is accepted only if it lowers the energy, i.e. if the
charge advances in the direction of ${\bf F}$.  Since only 
$3$ of the $6$ possible directions do so, the move is accepted with probability $1/2$.
If the work done by the force dominates the interaction energy (as it 
should for $F>>F_{c}$) then after one pass
a random half of all charges have moved forward.  On the next pass, a 
different random half move forward.  After several 
passes, the charges are completely 
disordered.  Although this argument assumed $F>>F_{c}$, we find 
that at $T=0$ the charges melt to a liquid at all $F\ge F_{c}$.
The randomness of choosing proposed moves thus has a 
dramatic effect on the steady state order.  In contrast, in CTMC, 
moves are chosen according to a probabilistic distribution which 
sharpens dramatically as $T$ decreases.  
Unlikely moves are almost never chosen, and favorable moves 
are almost always chosen. The result, shown in Fig.\,\ref{f1}, is more 
structure in the moving steady states.  The rest of this work, 
therefore, focuses on CTMC.

For CTMC, the results of Fig.\,\ref{f1} show a discontinuous melting
transition at $T_{\rm m}=0.00325$.  In Fig.\,\ref{f2}b$-$d we show
representative plots of $S({\bf k})$ above and below $T_{\rm m}$.
We first consider $T>T_{\rm m}$.  Although the plot of 
$S({\bf k})$ at $T=0.004$ in Fig.\,\ref{f2}c shows sharp peaks at the
reciprocal lattice vectors of the ordered charge lattice,the
magnitude of these peaks is greatly reduced from those at $T<T_{\rm
m}$ (see Fig.\,\ref{f2}b).  We have carried out simulations
for a larger system, $L_{2}=120$, $L_{1}=500$ and
found $S({\bf k})$ to remain unchanged.  This lack of finite size
dependence in $S({\bf k})$ indicates that the system is a liquid with
short ranged translational order.  The peaks in $S({\bf k})$
result from large but finite correlation lengths.  Similar behavior 
was seen in simulations of vortices in a square Josephson junction array with 
{\it random} pinning \cite{Dominguez}; the prominent peaks at the
transverse wavevectors along the $k_y$ axis led those authors to
denote this state as a ``short ranged smectic''.

\begin{figure}
\epsfxsize=7.5truecm
\epsfbox{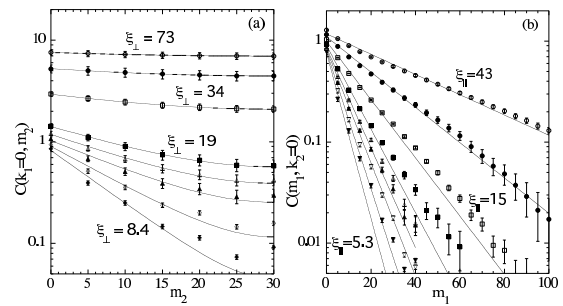}
\caption{(a) Transverse real space correlation $C(k_{1}=0,m_{2})$ vs.
$m_{2}$, and
(b) longitundinal real space correlation $C(m_{1},k_{2}=0)$ vs.
$m_{1}$.  Solid lines are fits to ${\rm e}^{-m/\xi}+{\rm
e}^{-(L-m)/\xi}$ and determine the correlation lengths $\xi_{\perp}$ 
and $\xi_{||}$.  Curves from top to bottom are for $T=0.00325, 0.0035, 
0.004, 0.0045, 0.005, 0.0055, 0.006, 0.007$; $F=0.1$. Representative
values for $\xi_\perp$ and $\xi_{||}$ are shown.
}
\label{f3}
\end{figure}

To estimate the correlations lengths we plot in 
Fig.\,\ref{f3}a,b $C(k_{1}=0,m_{2})$ vs. $m_{2}$
and $C(m_{1},k_{2}=0)$ vs. $m_{1}$; the first gives the decay of correlations in 
real space along $\hat a_{2}$ (averaged over the direction $\hat
a_{1}$), while the second gives the decay of correlations along $\hat
a_{1}$, parallel to the force ${\bf F}$ (averaged over $\hat a_{2}$). 
We plot only the values at integer multiples of the average particle
spacing $a_{v}=1/\sqrt f =5$; these define the upper envelop of
the damped oscillating correlations.
Fitting the data to the simple periodic decay $C\sim {\rm
e}^{-m/\xi}+{\rm e}^{-(L-m)/\xi}$ (where we use $L_{1}$ or $L_{2}$ as 
appropriate) gives the correlation lengths perpendicular, $\xi_{\perp}$,
and parallel, $\xi_{||}$, to the driving force ${\bf F}$.   
Representative values for $\xi_\perp$ and $\xi_{||}$ are shown in Fig.\,\ref{f3}
with $\xi_{\perp}\sim 2\xi_{||}$ near $T_{\rm m}$.
For Fig.\,\ref{f3}b our fit is only to points $m_{1}>\xi_{||}$. 
%
%
%

Although the liquid above $T_{\rm m}$ lacks translational order,
Fig.\,\ref{f1}b shows that hexatic orientational order
grows for $T<T^{*}\sim 0.007$.  Similar hexatic liquids have been
reported in continuum simulations with {\it random} pinning \cite{Ryu}.  In our case, the
periodic triangular grid breaks rotational symmetry and in principle 
gives finite hexatic order at any $T$.  It is unclear whether the
onset of growing hexatic order at $T^{*}$ is just this grid induced
effect, increasing as the correlations lengths grow larger than
the interparticle spacing, $\xi > a_{v}$, or whether it is a 
crossover remnant of what might be a true hexatic transition in
another geometry.

\begin{figure}
\epsfxsize=7.5truecm
\epsfbox{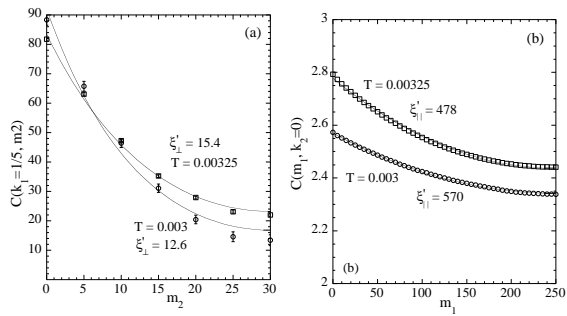}
\caption{(a) Transverse correlation between the smectic channels,
$C(k_{1}=1/5, m_{2})$ vs. $m_{2}$, and (b) longitudinal correlation
$C(m_{1},k_{2}=0)$ vs. $m_{1}$, in the smectic state for $T=0.003,
0.00325\le T_{\rm m}$, $F=0.1$.  Solid lines are fits to ${\rm e}^{-m/\xi}+{\rm
e}^{-(L-m)/\xi}$ and determine the correlation lengths
$\xi_{\perp}^{\prime}$ 
and $\xi_{||}^{\prime}$.
}
\label{f5}
\end{figure}

Next we consider $T<T_{\rm m}$.
$S({\bf k})$ for $T=0.003<T_{\rm m}$ is shown in Fig.\,\ref{f2}b.
First we note that the peaks at ${\bf K}$ on the $k_{y}$ axis 
(at  $k_{1}=0, k_{2}=1/5,2/5$) are sharp $\delta$-function Bragg 
peaks of height $S({\bf K})\simeq N_{c}=1200$.
We have computed the height of these peaks for smaller size systems and
find that they scale $\sim N_{c}$.  This indicates that this state
has long range smectic order; the particles are confined to
periodically spaced channels parallel to the driving force.
Next, we note that the peaks at finite $k_{x}$ (at $k_{1}=1/5,2/5$) are
essentially $\delta$-functions in $k_{x}$.  Simulations for
smaller size systems show that the height of these peaks scale
$\sim L_{1}$.  This indicates that the particles are perfectly ordered
within each smectic channel.  The finite width of these peaks, as $k_{y}$ 
varies, indicates that the ordered channels are randomly displaced
with respect to each other, with a finite correlation length
$\xi_{\perp}^{\prime}$.  To determine this correlation length we plot
in Fig.\,\ref{f5}a $C(k_{1}=1/5,m_{2})$ vs. $m_{2}$ and fit to a
periodic decay as earlier.  
Finally we return to the issue of the longitudinal order in the
smectic state.  Since the transverse order among the smectic channels
is short ranged, with finite $\xi_{\perp}^{\prime}$, we can regard
the channels as decoupled one dimensional systems.
Thus only short ranged
longitudinal order should be expected.  We have investigated this
issue by carrying out detailed finite size analyses on smaller size
systems.  We conclude that the smectic does in fact have a finite
longitudinal correlation length $\xi_{||}^{\prime}$, but that 
$\xi_{||}^{\prime}\gtrsim L_{1}$;  we find that when the system length is increased so that
$\xi_{||}^{\prime}\alt L_{1}$, the smectic becomes unstable to the liquid.
In Fig.\,\ref{f5}b we plot the longitudinal correlation
$C(m_{1},k_{2}=0)$ vs. $m_{1}$ for $T=0.003,0.00325$ just below
$T_{\rm m}$. Fitting to the periodic
exponential decay assumed earlier, we find $\xi_{||}^{\prime}\sim
500$.  For a smaller length, $L_{1}=120$, the smectic persisted up to
the higher $T=0.006$.  Our desire to supress $T_{\rm m}$ to low
temperatures, so as to see growing correlations in the liquid, was the 
reason we chose $L_{1}=500$ for the simulations reported on here.
While the smectic thus disappears in the true thermodynamic limit,
since $\xi_{||}^{\prime}$ grows exponentially as $T$ decreases, the
smectic will ultimately appear in a finite sized system at
sufficiently low $T$.  We find that once $\xi_{||}^{\prime}\gtrsim L_{1}$, 
the smectic is the stable state of the system; for
$L_{1}=L_{2}=120$, we have suceeded in cooling into the
smectic from the disordered liquid.  Our observation of a smectic state
on finite length scales, which becomes unstable to a liquid on large length scales,
agrees in part with arguments in Ref. \cite{Balents}.

Interacting 2D vortices in a periodic potential at finite
temperature have been simulated by several others using {\it continuum} 
dynamics.  The molecular dynamic simulations of Reichhardt and 
Zim\'anyi \cite{Reichhardt} and  Carneiro \cite{Carneiro} used square
periodic pins embedded in a flat continuum, with a number
of vortices equal to, or greater than, the number of pins.  Such models
cannot be well described by our discrete grid.
Closer to our model is that of Marconi and Dom\'{\i}nguez \cite{Marconi}
who simulate the RSJ dynamics of a vortex density $f=1/25$ in a {\it square} Josephson
array, and find an ordered moving vortex lattice.  
However in their case, the energy to move a single vortex forward
from its ground state position is $\Delta E_1\simeq 0.34$, whereas the energy barrier
between cells of the array is $U_0\simeq 0.12$.  The parameters of our
simulations, which assume $\Delta E_1<F<U_0$ (see discussion preceding Eq.\,(\ref{eS})),
are therefore in a more strongly pinned limit outside the range of their model.  It therefore
remains for future investigation to test if the results of the CTMC method
agree with that of continuum models in the corresponding limit.

We wish to thank D.~Dom\'{\i}nguez, M.~C.~Marchetti and A.~A.~Middleton
for helpful discussions.  This work was supported by DOE grant DE-FG02-89ER14017
and NSF grant PHY-9987413.


%


\end{document}